\documentclass[aps,reprint,prb,showpacs,amsmath,amssymb,citeautoscript,numerical]{revtex4-1}

\usepackage{graphicx} 
\usepackage{bm}
\setcitestyle{super}

\begin{document}

\title{Threshold of Terahertz Population Inversion and Negative 
Dynamic Conductivity in Graphene Under Pulse Photoexcitation}

\author{A.~Satou}\email{a-satou@riec.tohoku.ac.jp}
\author{V.~Ryzhii}
\author{Y.~Kurita}
\author{T.~Otsuji}
\affiliation{Research Institute of Electrical Communication, 
  Tohoku University, 2-1-1 Katahira, Aoba-ku, Sendai 980-8577, Japan}

\date{October 29, 2012}

\begin{abstract}
  We theoretically study the population inversion and negative
  dynamic conductivity in intrinsic graphene in the terahertz (THz)
  frequency range upon pulse photoexcitation with near-/mid-infrared 
  wavelength.
  The threshold pulse energy required for the population inversion
  and negative dynamic conductivity can be orders-of-magnitude lower 
  when the pulse photon energy is lower, due to the inverse
  proportionality of the photoexcited carrier concentration 
  to the pulse photon energy and to the weaker carrier heating.
  We also investigate the dependence of the dynamic conductivity on 
  the momentum relaxation time. The negative 
  dynamic conductivity takes place either in high- or low-quality 
  graphene, where the Drude absorption by carriers in the
  THz frequency is weak.
\end{abstract}

\maketitle
\newpage

\section{Introduction}
Graphene has attracted much attention for wide variety of device 
applications due to its exceptional electronic and optical 
properties~\cite{Novoselov-Nature-2005, Geim-PT-2007,
Bonaccorso-NaturePh-2010}.
Especially, THz devices such as lasers~\cite{Ryzhii-JAP-2007,
  Ryzhii-PSS-2008,Ryzhii-JAP-2009,Satou-JJAP-2011,Ryzhii-JJAP-2011,
  Ryzhii-JAP-2011} and 
photodetectors~\cite{Ryzhii-APEX-2008,Ryzhii-OR-2012,
  Vicarelli-NatureMat-2012} utilizing its high carrier 
mobility and gapless dispersion have been investigated (for a review
of these devices, see Ref.~\onlinecite{Otsuji-JPD-2012}).
In Refs.~\onlinecite{Ryzhii-JAP-2007,Ryzhii-PSS-2008,Ryzhii-JAP-2009,
  Satou-JJAP-2011,Ryzhii-JJAP-2011},
 we demonstrated that the population inversion can occur 
in optically pumped graphene at THz/far-infrared range of frequency 
and hence the lasing at such a range is possible, utilizing the 
gapless linear energy spectrum and relatively high optical-phonon
(OP) energy in graphene. 
Due to the energy spectrum $\varepsilon = \pm v_{_{F}}\hbar k$,
where $v_{_{F}} \simeq 10^{8}$ cm/s is the Fermi
velocity and $k$ is the wavenumber, the Fermi energy 
$\varepsilon_{_{F}}$ in 
equilibrium in intrinsic graphene is equal to zero. 
Hence, the electron and hole 
distribution functions at the bottom of the conduction band 
and the top of the valence 
band have values $f_{\text{e}}(0) = f_{\text{h}}(0) = 1/2$. This 
implies that upon photoexcitation one might expect to make 
the values of the distribution functions 
at low energies greater than one half, i.e., 
$f_{\text{e}}(\varepsilon) = f_{\text{h}}(\varepsilon) > 1/2$, 
corresponding to the population inversion. Such a 
population inversion leads to the total dynamic conductivity 
in graphene at THz/far-infrared frequencies being negative,
and lasing in graphene at such frequencies is possible.

Recently, we measured the carrier relaxation and recombination
dynamics in optically pumped epitaxial graphene on 
silicon~\cite{Karasawa-JIMTW-2010} and exfoliated 
graphene~\cite{Tombet-PRB-2012} using
THz time-domain spectroscopy based on an optical pump/THz\&optical
probe technique, and observed the amplification of THz radiation
by stimulated emission from graphene under pulse excitation.
To our knowledge, those are the first observation of the 
THz amplification using optically pumped graphene.
Those results demonstrate the possibility of realizing 
THz lasers based on graphene.

The carrier dynamics in optically pumped graphene strongly depends on
the initial temperature of carriers and the intensity of the 
optical pumping. With sufficiently low carrier concentration, 
i.e., at low temperature and upon weak pumping, photoexcited 
carriers are effectively accumulated
near the Dirac point via cascade emission of OPs, and
the achievement of the population inversion is expected to be
efficient~\cite{Ryzhii-JAP-2007,Ryzhii-PSS-2008,Ryzhii-JJAP-2011}.
On the contrary, at room temperature or upon not so weak pumping 
where the carrier concentration is rather high, the carrier-carrier 
(CC) scattering 
plays a crucial role in the dynamics after the pulse
excitation, due to fast quasi-equilibration of 
carriers.
In fact, ultrafast optical 
pump-probe spectroscopy on graphene has indicated that the 
quasi-equilibration by the CC scattering takes place within the 
time scale of 10 fs~\cite{Sun-PRL-2008, Dawlaty-APL-2008, 
Breusing-PRL-2009, Dani-PRB-2012},
which is much faster than the process of single OP emission.
In such a case, the pulse excitation makes carriers very hot
initially, and the energy relaxation and recombination 
via OP emission follow.

Previously, we studied the population inversion and the negative
dynamic conductivity in intrinsic graphene at room temperature 
upon the pulse excitation
with photon energy $0.8$ eV, corresponding to the wavelength 
$1.55$ $\mu$m, and showed that they can be achieved with the pulse
energy fluence above the 
threshold~\cite{Satou-JJAP-2011, Satou-PIERS-2012}.

In this paper, we investigate the dependence of the threshold
pulse fluence for the THz negative dynamic conductivity on 
the pulse photon energy, using the model extended from that 
developed in Refs.~\onlinecite{Satou-JJAP-2011, Satou-PIERS-2012} to
take into account the effect of the Pauli blocking.
We also study the dependence of the dynamic conductivity on the
momentum relaxation time $\tau$. We consider the two limiting cases
where $\omega\tau \geq 1$, corresponding to high-quality graphene,
and where $\omega\tau\ll1$, corresponding to low-quality graphene.

\section{Equations of The Model}

Assuming the quasi-Fermi distribution of carriers due to the 
quasi-equilibration by the CC scattering, i.e.,
$f_{\varepsilon} = \{1+\exp[(\varepsilon-\varepsilon_{F})/k_{B}T_{c}]\}^{-1}$,
the time evolution of the distribution is represented through the
quasi-Fermi level $\varepsilon_{F}(t)$ and the carrier temperature 
$T_{c}(t)$. Here, both electron and hole distributions can be
expressed by a single ``carrier distribution'' because we consider
intrinsic graphene; due to the symmetric dispersion for electrons and
holes in the energy range under consideration, i.e., 
$\varepsilon < 1$ eV, their distributions remain identical even upon 
the pulse excitation. The rate equations which determine the
quasi-Fermi level and carrier temperature can be obtained from the
quasi-classical Boltzmann 
equation~\cite{Satou-JJAP-2011, Satou-PIERS-2012, Knorr-papers}:
\begin{eqnarray}\label{EqRateConc}
  \frac{d n}{dt} &=& \frac{2}{\pi}
  \sum_{i=\Gamma, \text{K}} \int_{0}^{\infty}kdk\times
  \nonumber\\
  & & \left[
    \frac{(1-f_{\hbar\omega_{i}-\varepsilon})(1-f_{\varepsilon})}
    {\tau_{i,\rm inter}^{(+)}}
    -\frac{f_{\varepsilon}f_{\hbar\omega_{i}-\varepsilon}}
    {\tau_{i,\rm inter}^{(-)}}\right],
\end{eqnarray}
\begin{eqnarray}\label{EqRateEnergy}
  \frac{d{\cal E}}{dt} &=& \frac{2}{\pi}
  \sum_{i=\Gamma, \text{K}} \int_{0}^{\infty}kdk\times
  \nonumber \\
  & &\left\{\hbar\omega_{i}\left[
    \frac{f_{\varepsilon}(1-f_{\varepsilon+\hbar\omega_{i}})}
    {\tau_{i,\rm intra}^{(+)}}
    -\frac{f_{\varepsilon}(1-f_{\varepsilon-\hbar\omega_{i}})}
    {\tau_{i,\rm intra}^{(-)}}
    \right]\right.
  \nonumber \\
  & & \left.+\varepsilon\left[
    \frac{(1-f_{\hbar\omega_{i}-\varepsilon})(1-f_{\varepsilon})}
    {\tau_{i,\rm inter}^{(+)}}
    -\frac{f_{\varepsilon}f_{\hbar\omega_{i}-\varepsilon}}
    {\tau_{i,\rm inter}^{(-)}}
    \right]\right\},
\end{eqnarray}
where $n=n(\varepsilon_{F}, T_{c})$ and 
${\cal E}={\cal E}(\varepsilon_{F}, T_{c})$ 
are the concentration and energy density of either type of carriers,
which are found by integrating over $\bm{k}$ the distribution 
function multiplied by proper factors, the index $i$ runs over
types of OPs (long-wavelength $\Gamma$-OP and short-wavelength 
$K$-OP), and $\tau_{i,\rm intra}^{(\pm)}$ and 
$\tau_{i,\rm inter}^{(\pm)}$ are the intraband and interband 
scattering rates for OPs ($(+)$ for absorption and $(-)$ for 
emission). Those rates are on the order of sub-picoseconds
for high-energy carriers.

Note that the first term (intraband transition) 
in the curly brackets in 
Eq.~(\ref{EqRateEnergy}) contains the factor $\hbar\omega_{i}$, 
whereas the second term (interband transition) contains 
$\varepsilon$. These are energy that a carrier looses/acquires
in their corresponding transitions. The energy conservation
for the latter transition is implicitly satisfied because 
the rest of the OP energy $\hbar\omega_{i}-\varepsilon$ is
taken into account in the same equation for the other kind
of carrier. 
Here, we assume the equilibrium OPs and neglect the
nonequilibrium population of OPs due to their emission via 
OP scattering of carriers. The effects of the latter
have been discussed elsewhere~\cite{Ryzhii-JJAP-2011}.

Equations~(\ref{EqRateConc}) and (\ref{EqRateEnergy}) 
together with initial conditions form
a nonlinear system of equations for $\varepsilon_{_{F}}(t)$ and
$T_{c}(t)$, which can be solved numerically.
Equations~(\ref{EqRateConc}) and (\ref{EqRateEnergy}) are 
accompanied with the initial carrier concentration and energy density
from which the initial quasi-Fermi level and carrier temperature can
be found:
\begin{equation}\label{EqInitialCond}
  n|_{t=0} = n_{0}+\Delta n, \quad
  {\cal E}|_{t=0} = {\cal E}_{0}+\Delta {\cal E},
\end{equation}
where $n_{0}$ and ${\cal E}_{0}$ are the intrinsic carrier 
concentration and energy density, and $\Delta n$ and 
$\Delta {\cal E}$ are contributions of photogenerated carriers.
Those are equal to the concentration and energy density
generated from the pumping:
\begin{equation}\label{EqPhotogenerated}
  \Delta n \simeq 
    \frac{\pi\alpha \Delta J}{\sqrt{\epsilon}\hbar\Omega}
    (1-2f_{\hbar\Omega/2}|_{t=0}),\quad
    \displaystyle \Delta {\cal E} \simeq \frac{\hbar\Omega}{2}
    \Delta n,
\end{equation}
where $\alpha \sim 1/137$,
$\epsilon$ is the dielectric constant surrounding graphene,
$\hbar\Omega$ and $\Delta J$ are the pulse photon energy and the 
pulse fluence.

In Eq.~(\ref{EqPhotogenerated}), the last factor
$1-2f_{\hbar\Omega/2}$ 
roughly takes into account the Pauli blocking of the photoabsorption,
and the distribution function in it is approximated by the quasi-Fermi
distribution after the photoabsorption with
$\varepsilon_{F}|_{t=0}$ and $T_{c}|_{t=0}$ calculated 
self-consistently from Eqs.~(\ref{EqInitialCond}) and 
(\ref{EqPhotogenerated}).
This approximation underestimates the effect of the Pauli blocking 
by the {\it nonequilibrium} photogenerated carriers when the pulse is
so intense that the rate of the photogeneration
is comparable to that of the quasi-equilibration by the 
carrier-carrier scattering. On the other hand, it
overestimates the effect by time-dependent increase in the
distribution function {\it during} the pulse excitation, somewhat
compensating the underestimation described above.
In addition, we assume the pulse has sufficiently short
width that the OP scattering during the
absorption of the pulse is not effective, i.e., 
$\Delta t \lesssim \tau_{i,\rm inter}^{(+)}(\hbar\Omega/2),
\tau_{i,\rm intra}^{(+)}(\hbar\Omega/2)$.

To study not only the gain acquired by the population inversion
but also the net gain taking into account the loss due to the
Drude absorption of electromagnetic waves by carriers in graphene,
we introduce the real part of dynamic 
conductivity~\cite{Falkovsky-EPJ-2007} as it is related to the
net gain:
\begin{eqnarray}\label{EqConductivity}
  \text{Re}\,\sigma_{\omega} &\simeq &
  \frac{e^{2}}{4\hbar}(1-2f_{\hbar\omega/2})
  \nonumber \\
  & & 
  +\frac{2e^{2}v_{F}^{2}}{\pi}\int_{0}^{\infty}dk k
  \frac{\tau}{1+(\omega\tau)^{2}}
  \left(-\frac{d f_{\varepsilon}}{d\varepsilon}
  \right),
\end{eqnarray}
where $\tau$ is the momentum relaxation time.
The first term in Eq.~(\ref{EqConductivity}) corresponds to the
interband contribution which depends on the frequency only through
the distribution function and which can be negative when the rate of
stimulated emission exceeds the rate of absorption, i.e., when 
the population inversion takes place. 
Since the distribution function
is the quasi-Fermi distribution, the condition of the population 
inversion for the photon energy $\hbar\omega$ is simply represented as
$\varepsilon_{F} > \hbar\omega/2$. On the other hand, the second
term corresponds to the intraband contribution (Drude conductivity)
which is always positive in the system under consideration.

When the real part of the dynamic conductivity becomes 
negative at some
frequency, the electromagnetic wave at that frequency passing through 
optically pumped graphene is amplified and it serves as a gain medium.
Note that from Eq.~(\ref{EqConductivity}) the
largest achievable negative value of the conductivity 
is $-e^{2}/4\hbar$, corresponding to the amplification of 
$(2.3/\sqrt{\epsilon})$ \% of
the incoming electromagnetic wave.

%
\begin{figure}[t]
  \begin{center}
    \includegraphics{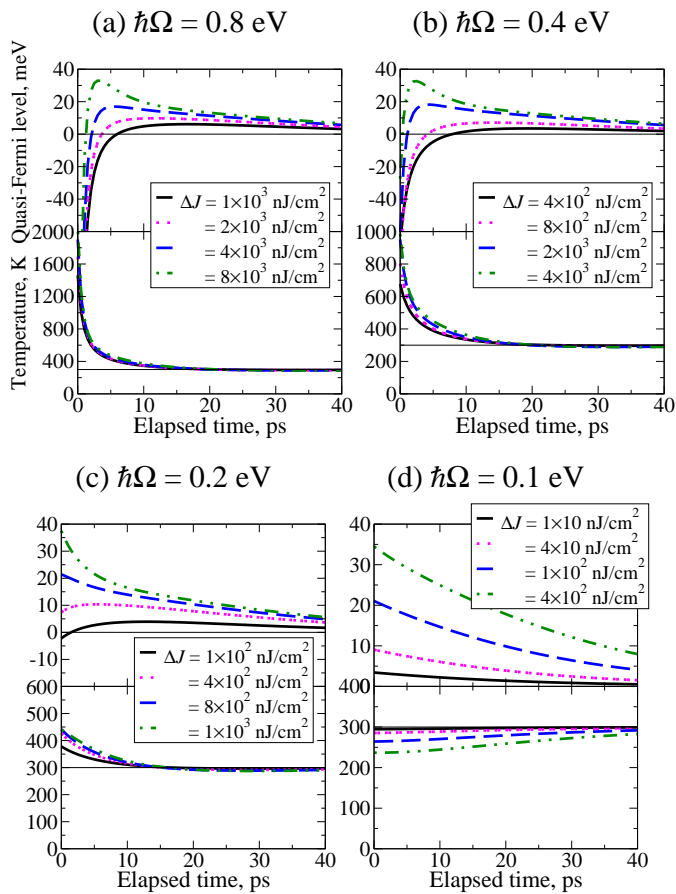}
  \end{center}
  \caption{\label{FigEpsTc}The time evolution of quasi-Fermi level
    and carrier temperature with different pulse fluence and pulse 
    photon energy [(a) $\hbar\Omega = 0.8$ eV, (b) 0.4 eV,
      (c) 0.2 eV, (d) 0.1 eV]. Room temperature (300 K) is indicated
    by thin solid lines.}
\end{figure}
\section{Results and Discussion}
\subsection{Time Evolution of Quasi-Fermi Level and Carrier 
Temperature}

Using the model above, we investigate the population inversion and
the time-dependent dynamic conductivity in intrinsic graphene
at room temperature after pulse excitation.
We varied the pulse photon energy $\hbar\Omega$ in the range
between $0.1-0.8$ eV corresponding to infrared wavelength of 
$12.4-1.55$ $\mu$m.
Besides, we set $\epsilon = 5.5$ corresponding to the effective 
dielectric constant of the interface between the air and SiC.
Here, we have used the pulse fluence rather than the pulse intensity
as a measure of pulse strength because the former does not depends
explicitly on the pulse photon energy, whereas the latter does
through the pulse width.

Figure~\ref{FigEpsTc} shows the time evolution of the quasi-Fermi
level and carrier temperature after the pulse excitation
with different pulse fluence and pulse photon energy.
It is clearly demonstrated in Fig.~\ref{FigEpsTc} that the
population inversion at THz frequencies, say, up to about
10 THz corresponding to $\hbar\omega/2 \sim 20$ meV is achieved
with some threshold pulse fluence,
and that it lasts on the order of 10 ps. 

The mechanism of the population inversion in the THz range
is explained as follows. Right after the pulse excitation,
the carrier temperature becomes very hot (except for the excitation
with low photon energy) and the quasi-Fermi level
can even be negative due to the quasi-equilibration of carriers 
with very large heat brought by the photogenerated carriers. 
After a few ps, carriers with high energy are relaxed via 
intraband OP emission and accumulated in the low energy region,
which is illustrated by the increase in the quasi-Fermi level.
As discussed in Ref.~\onlinecite{Satou-PIERS-2012}, 
the population inversion
lasts on the order of 10 ps due to the imbalance between
the time scales of intra- and inter-band OP emission.

\begin{figure}[t]
  \begin{center}
    \includegraphics{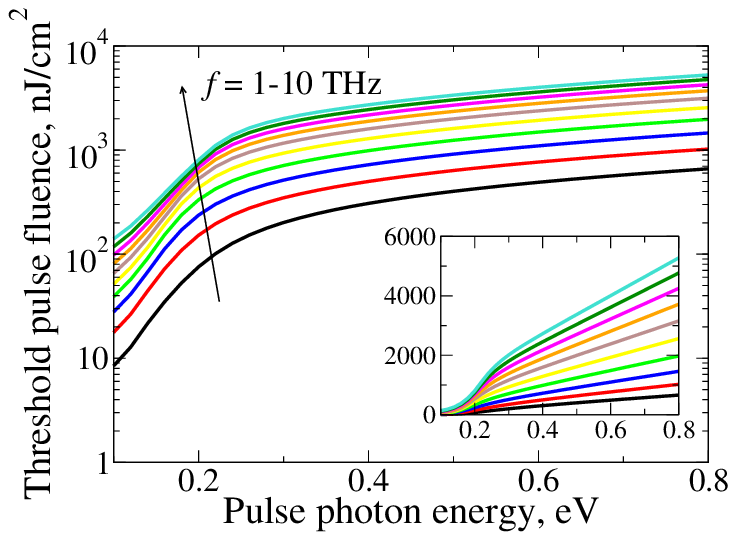}
  \end{center}
  \caption{\label{FigThreshold}The dependence of the threshold
    fluence of the population inversion on the pulse photon energy
    for different frequencies of THz photons. The inset shows
    the same dependence but with linear scale.}
\end{figure}
\begin{figure*}[t]
  \begin{center}
    \includegraphics{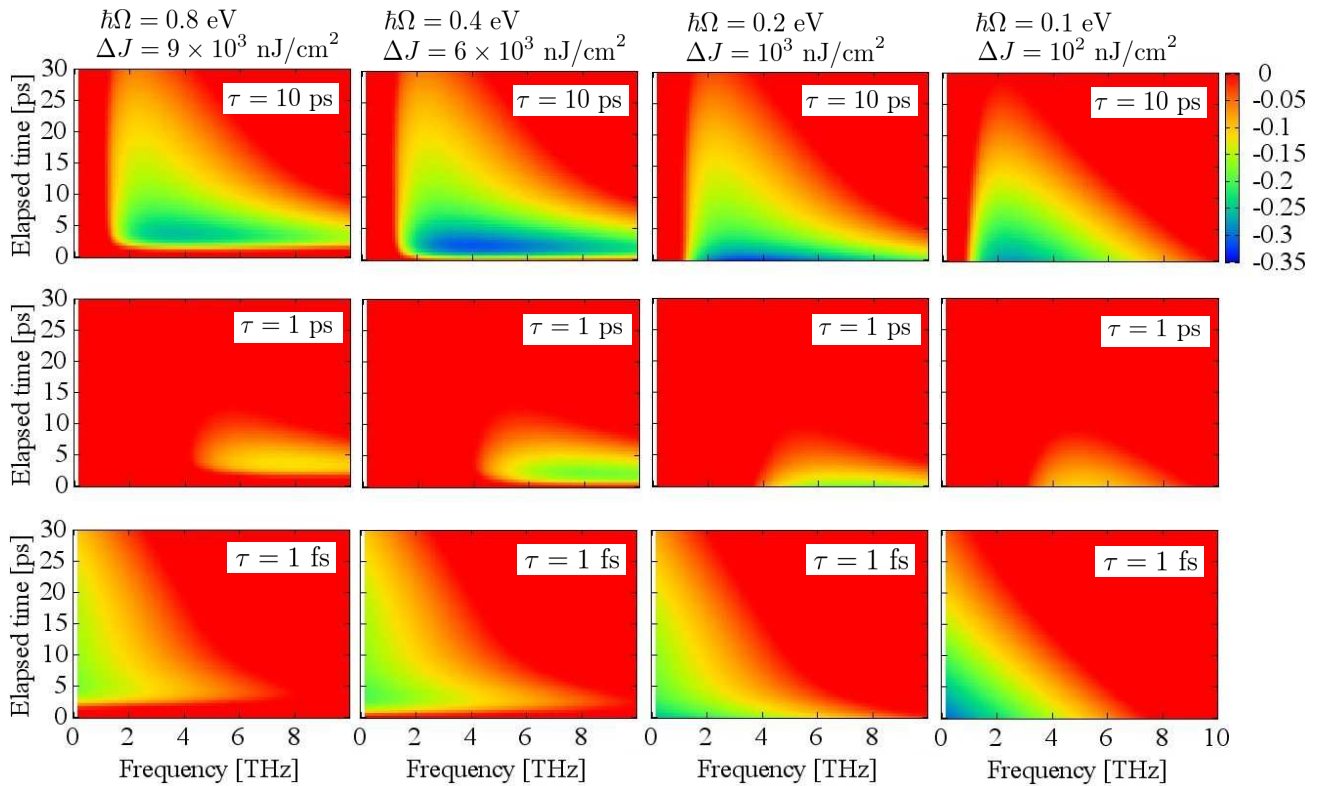}
  \end{center}
  \caption{\label{FigConductivity3D}The time-evolution of
  the dynamic conductivity in the THz range normalized to 
  $e^{2}/4\hbar$ and cut off above zero.
  The columns from left to right correspond to
  cases with $\hbar\Omega = 0.8$, $0.4$, $0.2$, and $0.1$ eV and
  $\Delta J = 6\times10^{3}$, $1\times10^{3}$, and $1\times10^{2}$
  nJ/cm$^{2}$, respectively, whereas the rows from top to bottom
  correspond to cases with $\tau = 10$ ps, $1$ ps, and $1$ fs, 
  respectively.}
\end{figure*}

Comparing Figs.~\ref{FigEpsTc}(a)-(d), one can see that the
threshold fluence of the population inversion becomes lower
as the pulse photon energy becomes lower.
This is also seen in Fig.~\ref{FigThreshold} that illustrates
the dependence of the threshold fluence of the population inversion
on the pulse photon energy for different frequencies of THz photons.
There are several factors for this lowering. First, 
the lower fluence is needed for the pulse with lower photon energy 
to have the same amount of carrier concentration [more precisely,
$\Delta J \propto \hbar\Omega$, as seen
in Eq.~(\ref{EqPhotogenerated})]. Second, the heating 
of carriers becomes less and the carrier temperature becomes lower
[even the cooling can occur as shown in
Fig.~\ref{FigEpsTc}(d); see below for more details],
resulting in their denser accumulation in the low energy region. 
Finally, the interband OP emission, i.e., the recombination
is suppressed due to the lower carrier temperature.
All these result in the nonlinear lowering of the threshold as
the pulse photon energy decreases; 
Fig.~\ref{FigThreshold} exhibits almost linear dependence
of the threshold in the region of large photon energy,
caused by the first factor, and
the sharp drop in the region of photon energy below $0.25$ eV,
caused by the second and third factors.

The effect of the Pauli blocking depends strongly on the pulse
photon energy.
For $\hbar\Omega=0.8$ eV, the changes in the quasi-Fermi level and 
carrier temperature were negligible for any pulse fluence
in consideration when the Pauli blocking was taken into account.
On the contrary, for $\hbar\Omega=0.1$ eV, 
the threshold fluence changed $1.3-1.8$ times larger depending
on the frequency, and the quasi-Fermi level achievable became
smaller.

As has been discussed in 
Refs.~\onlinecite{Ryzhii-PSS-2008, Ryzhii-JJAP-2011},
the cooling in the case of low pulse photon energy occurs
because the initial photoexcitation can give the lower energy to
carriers than in thermal equilibrium. In fact, 
the energy of photogenerated carriers, $\hbar\Omega/2$, is
lower than the average energy of carriers in intrinsic graphene 
at room temperature, ${\cal E}/n\approx 0.0567$ eV, when
$\hbar\Omega < 0.13$ eV. Besides, the cooling after the
heating by the photoexcitation (albeit small) can also occur
as seen in Figs.~\ref{FigEpsTc}(a)-(c). This is associated with
the discrete nature of OP emission/absorption which cannot
exactly bring the carrier temperature to the lattice temperature.
The latter is to be accomplished via the acoustic phonon scattering 
and/or radiative generation/recombination with substantially
longer time scale.

\subsection{Time-Evolution of Dynamic Conductivity in Cases of High-
  and Low-Quality Graphene}

\begin{figure*}[t]
  \begin{center}
    \includegraphics{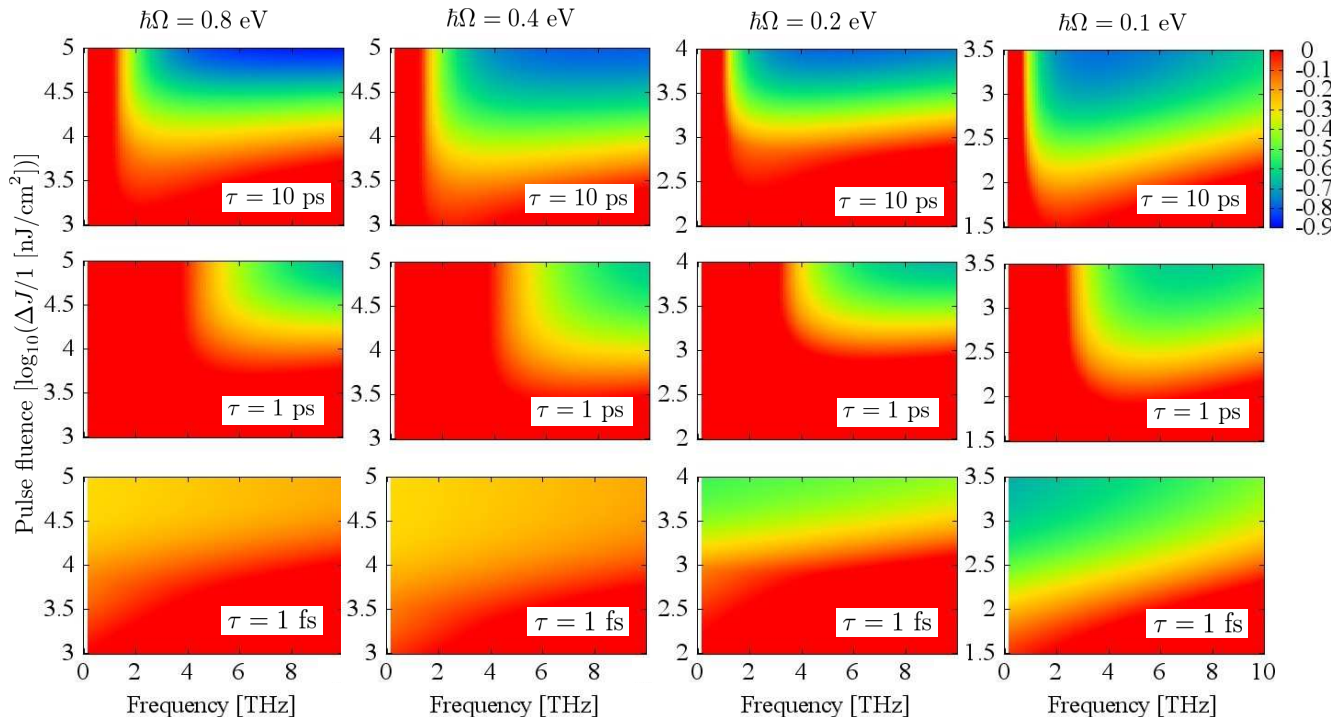}
  \end{center}
  \caption{\label{FigMaxConductivity3D}The maximum negative value
    of the dynamic conductivity in its time duration as a function
    of frequency and pulse fluence.
    The columns from left to right correspond to
    cases with $\hbar\Omega = 0.8$, $0.4$, $0.2$, and $0.1$ eV,
    respectively, whereas the rows from top to bottom correspond to
    cases with $\tau = 10$ ps, $1$ ps, and $1$ fs, respectively.}
\end{figure*}

The dynamic conductivity in Eq.~(\ref{EqConductivity}) depends 
primarily on the carrier 
distribution, i.e., the carrier concentration, as well as the 
momentum relaxation time $\tau$. 
Since the photoexcitation not only leads to the population inversion
but also to the increase in the carrier concentration, the
negative dynamic conductivity arises if the increase in the 
absolute value of the interband conductivity overcomes
the increase in the Drude absorption in Eq.~(\ref{EqConductivity}).
For the realization of low threshold 
of negative dynamic conductivity in the THz range, 
it is necessary to suppress the Drude absorption. 
This can be done by using either
good-quality graphene such that $\omega\tau \geq 1$ or
high-quality graphene such that $\omega\tau\ll1$. As can be seen
in Eq.~(\ref{EqConductivity}), in the former case, the high-frequency
``cut-off'' of the Drude conductivity occurs. In the latter case,
the Drude conductivity ceases in the entire frequency range.

Figure~\ref{FigConductivity3D} illustrates the time-evolution of 
normalized dynamic conductivity 
$\text{Re}\,\sigma_{\omega}/(e^{2}/4\hbar)$
in the THz range with different pulse
photon energy and momentum relaxation time. The pulse fluence
for each pulse photon energy was chosen so that sufficiently
large negative dynamic conductivity is obtained.
It demonstrates that the negative dynamic conductivity can be
achieved in the THz range with sufficiently high pulse fluence,
and that the same level of negative dynamic conductivity can be
achieved with much lower pulse fluence for lower pulse photon energy.
It also shows that its magnitude, frequency range, and time duration 
depend strongly on the momentum relaxation time.

For $\tau = 10$ ps, the negative dynamic conductivity occurs down
to $1$ THz and extends to more than $10$ THz, and it duration 
is $5-30$ ps depending on the frequency. The duration is longer for
lower frequency as is naturally understood from the relaxation of
the quasi-Fermi level shown in Fig.~\ref{FigEpsTc}.
For $\tau = 1$ ps, the negative dynamic conductivity occurs between 
$4$ and $10$ THz with smaller magnitude and much shorter duration.
This indicates that the high quality of graphene having
longer momentum relaxation time is preferable.
As can be seen in Fig.~\ref{FigConductivity3D}, 
the minimum frequency for the negative dynamic conductivity 
is primarily determined by the momentum relaxation time,
although it slightly shifts to the lower side in case of
the lower pulse photon energy.

On the contrary, for $\tau = 1$ fs, the negative dynamic conductivity
occurs in the low frequency region below $6$ THz.
Note, however, that the meaning of the ``time-dependent
dynamic conductivity'' is valid only if its variation is slower than
the frequency under consideration, approximately around or above 
a few hundred GHz in our case. This means we can have the negative 
dynamic conductivity down to a few hundred GHz.
The upper limit of the frequency range having the negative dynamic 
conductivity in this case is smaller than in the case where 
$\omega\tau\ll1$. This is because the Drude conductivity in this case
 is almost constant in the THz range, whereas it vanishes
due to the high-frequency ``cut-off'' in the latter case.

Our calculation showed that even for $\hbar\Omega = 0.1$ eV 
the threshold momentum relaxation time of the negative dynamic 
conductivity is
$\tau = 5$ fs, and above it the dynamic conductivity does not occur
in the THz range, no matter how large the pulse fluence is. 
This threshold roughly corresponds to the value 
at which the Drude term in Eq.~(\ref{EqConductivity}) at 
zero-frequency limit exceeds the largest negative value of the 
interband contribution, i.e., 
$\tau \sim (\pi/8\log2)(\hbar/k_{_{B}}T_{l})$.

Figure~\ref{FigMaxConductivity3D} shows the maximum negative value
of the dynamic conductivity in its time duration as a function
of frequency and pulse fluence with different pulse photon energy
and different momentum relaxation time.
It illustrates that the threshold pulse fluence as well as the pulse 
fluence needed to obtain the same value of the negative dynamic
conductivity are decreased in several orders of magnitude 
for any frequency as the pulse photon energy is decreased.
Comparing the cases with $\tau=10$ and $1$ ps, 
it is seen that
the decrease in the momentum relaxation time causes 
rather large increase in the minimum frequency for the negative 
dynamic conductivity and the increase in the threshold pulse fluence 
for frequency near the minimum, although the latter for frequency
sufficiently above the minimum is insignificant due to the 
high-frequency ``cut-off'' of the Drude conductivity.
As mentioned above already, the minimum frequency for the negative 
dynamic conductivity shifts to the lower side for the lower pulse 
photon energy. In fact, the frequency can be down to about $700$ GHz
with $\hbar\Omega = 0.1$ eV and $\tau = 10$ ps, whereas
it is around $1.1$ THz with $\hbar\Omega = 0.8$ eV and $\tau=10$ ps,
demonstrating the superiority of the pumping by lower photon energy.
For $\tau=1$ fs, the threshold pulse fluence for each photon energy
is on the same order in comparison with that for $\tau=10$ ps.

Also, it should be noted that the maximum saturates below the
``fundamental limit'' (i.e., $-e^{2}/4\hbar$) for very high 
pulse fluence. For example, for $\tau=10$ ps and $\hbar\Omega=0.1$
eV, it saturates with the value
$\text{Re}\,\sigma_{\omega}/(e^{2}/4\hbar) \approx -0.4$ 
at $f = 1$ THz and $\approx -0.8$ at $f=2$ THz.
The reason for this is that, for very high pulse fluence,
the increase in the 
Drude absorption due to the increase in the carrier concentration 
exceeds the interband contribution even though very high quasi-Fermi
level is obtained, and the maximum negative value of the dynamic
conductivity corresponding to the saturation value is reached
at later time after the relaxation of the carrier concentration.
The Pauli blocking also takes part in the saturation in cases of
lower photon energies.
The saturation values for $\tau=1$ fs are smaller than those 
for $\tau=10$ ps, especially, about 3 times smaller with 
$\hbar\Omega=0.4$ and $0.8$ eV. Again, this is due to the nonzero,
almost-constant value of the Drude conductivity in the THz range
in the limit $\omega\tau\ll1$.

\section{Conclusions}

We studied the population inversion and negative dynamic conductivity
in intrinsic graphene upon pulse photoexcitation with the pulse
photon energy in $0.1-0.8$ eV and their dependences on the pulse
fluence.
We showed that their threshold pulse fluence can be lowered
by two orders of magnitude with the lower pulse photon energy.
This is attributed to the inverse proportionality of the 
photogenerated carrier concentration to the photon energy, 
as well as to the weaker carrier heating
which results in the denser accumulation of carriers in the
low energy region and in the suppression of recombination via
interband OP emission.
We also showed that the negative dynamic conductivity takes place 
in the long relaxation time limit $\omega\tau \geq 1$, 
where the Drude absorption in the THz range becomes small by 
the high-frequency ``cut-off'', 
and in the very short relaxation time limit $\omega\tau\ll1$,
where it ceases in the entire frequency range.
We demonstrated that the pulse excitation by lower pulse photon
has both lower threshold pulse fluence and the lower
minimum frequency for the negative dynamic conductivity down to
about $700$ GHz with $\hbar\Omega = 0.1$ eV and $\tau = 10$ ps.

\begin{acknowledgments}
This work was supported by JSPS Grant-in-Aid for Specially Promoted
Research.
\end{acknowledgments}


\end{document}